\documentclass[12pt]{iopart}
\usepackage{graphicx}
\usepackage{here}
\def\bb{b\overline{b}}
\def\qq{q\overline{q}}

\def\mb{$\mu$b}
\def\pthat{\hat{P_{t}}}
\def\py{Pythia}

\begin{document}

\title[]{A study of $\bb$ 
production mechanisms in PYTHIA~\footnote{Talk presented at Phenomenology Workshop on Heavy Flavours 
and CP Violation, Durham UK}}

\author{Simone Gennai\dag   
\footnote[3]{E-mail: simone.gennai@pi.infn.it}
}

\address{\dag\ I.N.F.N. Pisa, Italy}

\begin{abstract}
The influence of the $\hat{P_{t}}$ cutoff on the production of $\bb$ events
in \py\ has been studied. Two samples of
events generated with  $\hat{P_{t}} \ge 1 GeV$ (default) and with
$\hat{P_{t}} \ge 10$ GeV are investigated.
In versions 5.75 and 6.115 of \py\ there are some critical differences 
between the two
samples for what concerns contribution in $\bb$ production mechanisms.
Version 6.152, taken as the CMS official one, has been tested also.
\end{abstract}




\section{Introduction}
CMS (Compact Muon Solenoid) is a general purpose experiment under construction
at CERN. The description of the apparatus is given elsewhere \cite{intro:cms}. 
CMS will be built on LHC (Large Hadron Collider) a $pp$ collider with
$\sqrt{s}$=14 TeV and a maximum luminosity of $10^{34}\,cm^{-2}\,s^{-1}$
\cite{intro:lhc}~.
One of the main physics topics of the Collaboration will be the study of the B hadron physics.
In fact apart from the interest ``per se'' of this topic, 
b-jets are also the main background for several
other physics channels~\cite{intro:acuto}, \cite{intro:btau}.
In this article, I present a study on $\bb$ production mechanisms
at LHC energies using the Monte Carlo package \py~\cite{intro:pyt} as event generator.\\
At a hadron collider  $\bb$ pairs  are produced by the following mechanisms:
\begin{enumerate}
\item {\bf Pair production} (gluon or quark fusion): $gg \rightarrow \bb$, 
$ \qq\rightarrow \bb$. For each event a $\bb$ pair is produced in the hard interaction.
\item{\bf Flavour excitation}: $bq \rightarrow bq$, $ bg\rightarrow bg$.
 A heavy flavour from the parton distribution of
one beam particle is put on mass shell by
scattering  against a parton of the other beam. As  $b$ quark is not a valence flavour
 it must come from a branching $g \rightarrow \bb$.
\item  {\bf Gluon splitting}:  $g \rightarrow \bb$. 
It occurs in the initial- or final-state shower but no heavy flavours
 are produced in the hard scattering.
\end{enumerate}
A more detailed definition of the mechanisms can be found in ref~\cite{intro:hep}.
As these mechanisms give rise to different kinematic configurations of the final state is important to understand
their  contribution to the total cross section.\\
The only way to generate (i), (ii) and (iii) in \py\ is using the steering card 
MSEL=1, which generates QCD $2 \rightarrow 2$ events.
In order to have higher  statistics, and save computation time the user 
can set kinematic cuts at generation level which limits the available portion of phase
 space~\cite{intro:trig}, 
increasing the 
efficiency of selection cuts.
One of this cuts is the $\pthat$ cutoff.~\footnote{$\pthat$ is defined as the transverse momentum of the outgoing partons in the 
center of mass system of the ingoing partons.}\\ 
In the following sections I will show  differences between the use of the 
\py\ default cutoff (1 GeV) and a  10 GeV  cutoff. Three versions of the generator  have been considered: 
5.75, 6.115 and 6.152 which has been chosen as event generator for the next CMS physics studies.
\section{$\bb$ production cross sections}
Two samples have been prepared with the three versions of \py\  to investigate the influence of the
$\pthat$ cutoff on the production of $\bb$ events.  Both of them have
been generated using MSEL=1 and contain events with only one $\bb$ pair.
The first sample (SAMPLE~A) has been generated
with the default $\pthat$ cutoff (1 GeV). The second sample
(SAMPLE B) has been generated with $\pthat \ge 10$ GeV.
For both samples the $\bb$ production cross section for  $\pthat \ge 10$~GeV has been computed
using the following formula:
\begin{equation}
\sigma_{ \bb }^{(\pthat\ge 10~GeV)}  =  \frac{N_{b\overline{b}}^{(\pthat\ge 10~GeV)}}{N_{tot}}\cdot \sigma_{tot}
\end{equation}
where $N_{b\overline{b}}^{(\pthat\ge 10~GeV)}$ is the number of $\bb$ events with $\pthat \ge$ 10 GeV,
 $N^{tot}$ is the total number of events
generated and $\sigma^{tot}$ is the total (QCD $2 \rightarrow 2$)
cross section given by \py\ . The results of the computation for the three versions of the package 
are shown in Tab.\ref{tab:res1} (the values of the cross section are given in \mb).\\
\begin{table}[h]
\begin{center}
\begin{tabular}{|l|l|l|l|l|l|l|} \hline\hline 
cross section ($\pthat \ge$ 10 GeV) ($\mu$b)& \multicolumn{2}{|c|}{P 5.75}&
\multicolumn{2}{|c|}{P 6.115}&\multicolumn{2}{|c|}{P 6.152} \\ \hline
 & A &B & A &B & A &B \\ \hline
$\sigma_{\mbox{{ \rm pair production}}}$ & {\bf 20}&   {\bf 20} &  {\bf 47} &  {\bf 51} & {\bf 48} & {\bf 52} \\ \hline
$\sigma_{\mbox{{\rm flavour\,\, excitation}}}$ & {\bf -} &   {\bf 110} &{\bf  -} &   {\bf 221} & {\bf 211} & {\bf 236}  \\
\hline
$\sigma_{\mbox{{\rm gluon splitting}}}$ &  {\bf 130} &    {\bf 130}  & {\bf 173} &  {\bf 193}  & {\bf 159} & {\bf 173}\\ 
\hline 
$\sigma_{\bb}$ & {\bf 150} & {\bf 260} & {\bf 220} & {\bf 465} & {\bf 418} & {\bf 461}  \\ \hline \hline
\end{tabular}
\end{center}
\caption{\label{tab:res1} Contributions to the  $\bb$ production cross section 
for $\pthat \ge 10$~GeV with \py\ 5.75, 6.115 and 6.152 (in \mb).}
\end{table}

All double countings have been avoided requiring the number of b-quarks coming from the hard interaction to be 2 
(pair production), 1 (flavour excitation) or 0 (gluon splitting).
As it appears clear from the table for version 5.75 and 6.115 the cross section is $\pthat$ dependent.
The problem is related to flavour excitation which in this two versions cannot be generated with default $\pthat$ 
cutoff (sample A). The difference in the $\sigma_{\bb}$ (for sample A) between 5.75 and 6.115 is due to the different choice of
the default parton distribution function.
The interpretation of the observed phenomena, given by T.~Sj\"ostrand and
E.~Norrbin, is reported in \cite{lhcsmws}.
In version 6.152 the problem related to flavour excitation has been set but there is still a 
difference of about 10\% between the cross sections of the two samples 
(for all versions statistical errors can be neglected). \\
The difference in the value of cross sections is not
very important, because the results are usually normalized to the
total $\bb$ cross section of 500~$\mu$b. On the other hand, the
different contributions for the production mechanisms could be more dangerous, as they can lead
to different kinematic distributions, and therefore affect the
efficiencies for different selections.\\
Fig. \ref{fig:cross} illustrates the difference in the $\bb$
production cross sections due to the additional contribution of the
flavour excitation mechanism in sample~B.\\

\begin{figure}[ht]
\begin{center}
\resizebox{8cm}{8cm}{\includegraphics{sigmahard.epsi}}
\caption{Cross section with the two different cutoff on $\hat{P_{t}}$.
The solid line is with the default cutoff (1 GeV).
The picture is obtained with \py\ 5.75.}
\label{fig:cross}
\end{center}
\end{figure}
\section{Kinematic distributions}
\begin{figure}[ht]
\begin{center}
\resizebox{8cm}{5cm}{\includegraphics{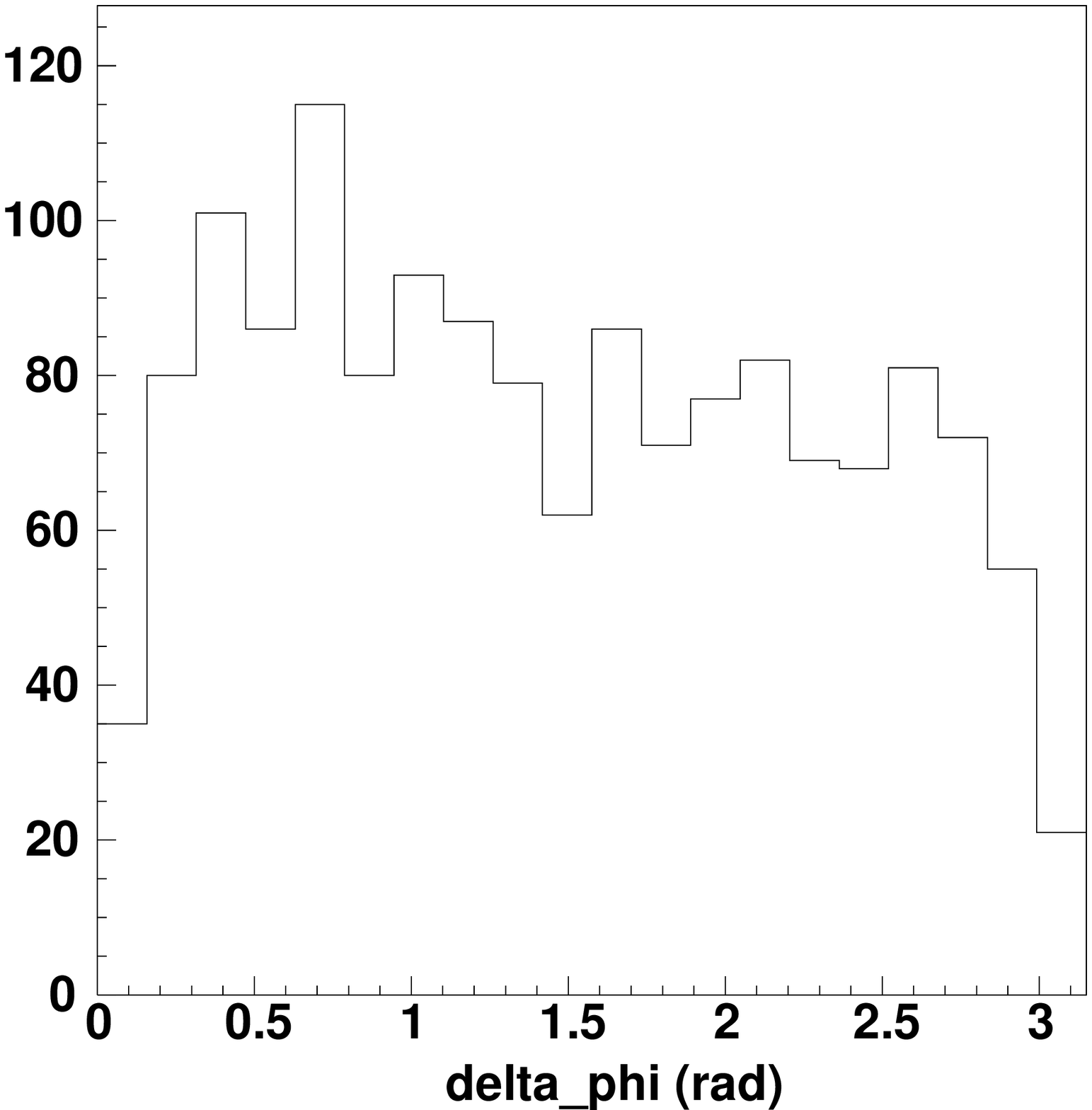}}
\resizebox{8cm}{5cm}{\includegraphics{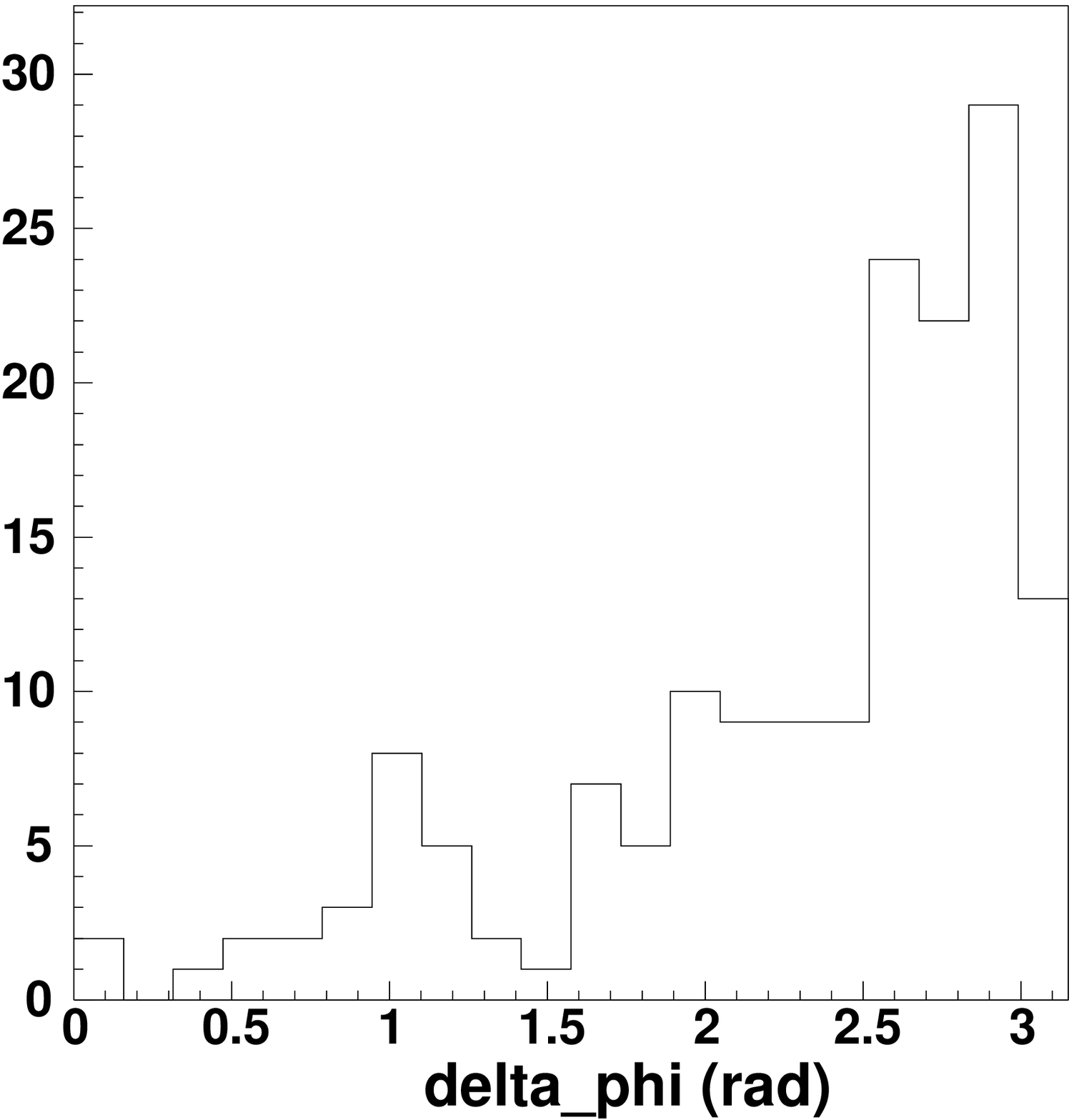}}
\resizebox{8cm}{5cm}{\includegraphics{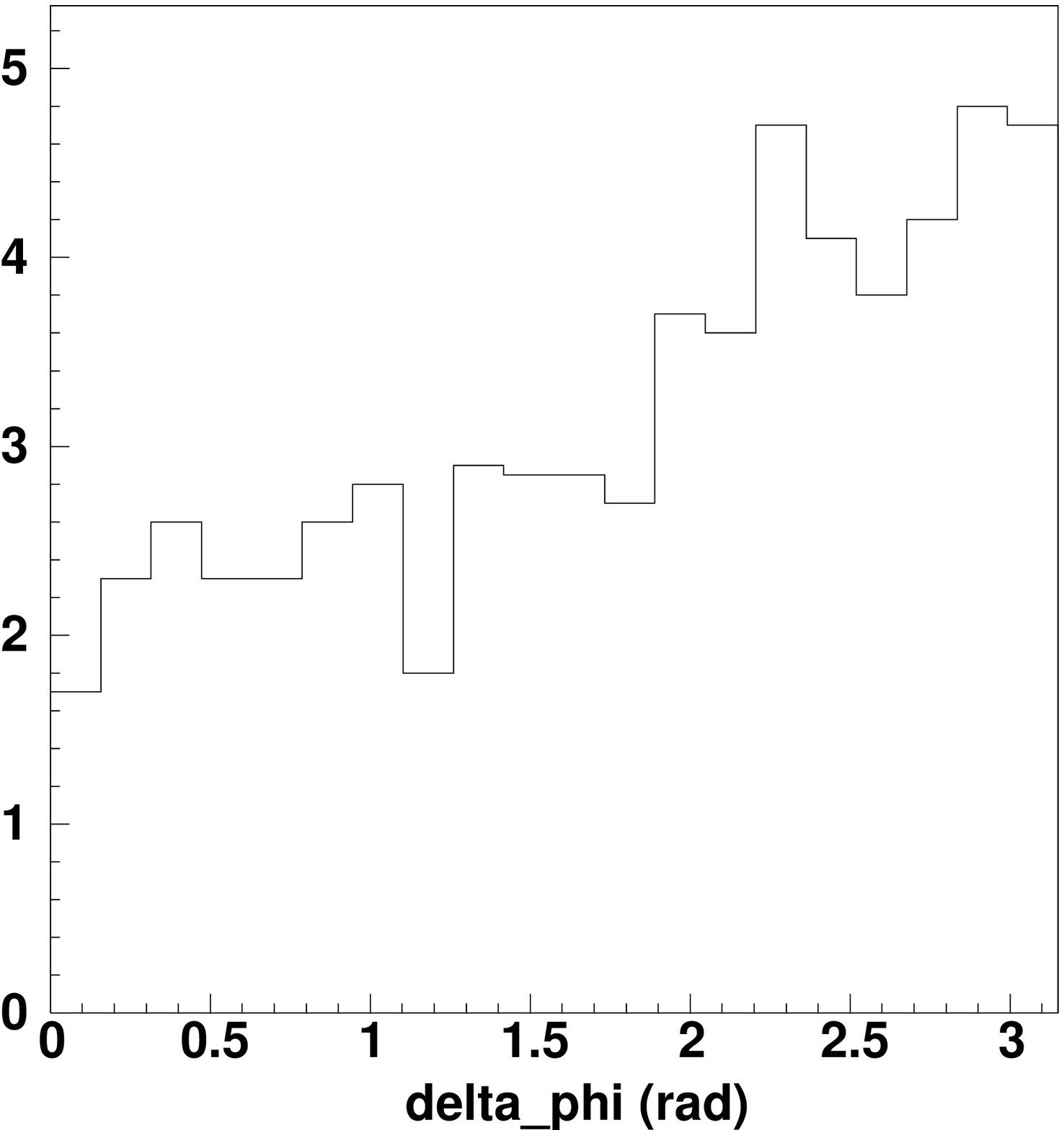}}
\caption
{Distribution of the angle between the two $b$ quarks in the transverse plane: the upper one
is gluon splitting, the middle is gluon fusion, the last is related to
flavour excitation. The plots are in arbitrary units.}
\label{fig:distr}
\end{center}
\end{figure}

The main kinematic parameters which define the signature of $\bb$
event are the transverse momenta and pseudorapidities of the $b$
quarks, and the angular distance $\Delta\phi$ between their directions in the
transverse plane. The first two parameters have similar distributions
in both samples. The $\Delta\phi$ distribution is shown in
fig. \ref{fig:distr} for the three different  mechanisms.
For what concerns gluon splitting, the distribution is slightly peaked
at small $\Delta\phi$. The angle between the two $b$-quarks produced
by the gluon-fusion mechanism has a peak at $\Delta \phi \sim \pi$, as
expected, since in the process $gg \rightarrow \bb$ the b-quarks are
produced back-to-back in the transverse plane.  The last distribution corresponds to the flavour excitation production
mechanism, for which the back-to-back topology is preferred. We can conclude that the total $\Delta \phi$ distributions of
sample A and sample B are slightly different. Some care should be taken about this, as it could affect the estimated
efficiency of selection cuts.

\section{Conclusions}
Versions 5.75 and 6.115 show a big discrepancy in the total $\bb$ production cross section when 
using two different cutoffs on $\hat{P_{t}}$. 
The problem is due to flavour excitation which can be generated only in sample B but gluon splitting and 
pair production are stable and show the same cross section in the two samples.
In both versions the use of a different $\hat{P_{t}}$.
cutoff from the default one leads to slightly different $\Delta
\phi_{\bb}$ distributions caused by the presence of flavour  excitation. \\
In version 6.152 everything is set, a part a little inconsistency of about 10\% on which I am working 
in order to understand the cause.\\

\section{Acknowledgements}
I really thank Andrei Starodumov, Fabrizio Palla and Roberto Dell'Orso who helped me to make this study.\\
I also would like to thank M.~Mangano, P.~Nason and  G.~Ridolfi for their collaboration and fruitful discussions.
Special thanks to T.~Sj\"ostrand and E.~Norrbin for detailed explanations
of observed \py\ behaviour and recomendations for the generation of $\bb$ events.

\section*{references}

\end{document}